\documentclass[aps,prb,twocolumn,a4paper,floatfix,showkeys,superscriptaddress]{revtex4}

\usepackage{amssymb}
\usepackage{amsmath}
\usepackage{graphicx}
\usepackage{color}
\usepackage{longtable}

\begin{document}
\topmargin +3mm

\title{Electronic and magnetic properties of the interface LaAlO$_3$/TiO$_2$-anatase from density functional theory}

\author{Mariana Weissmann}
\email{weissman@cnea.gov.ar}

\author{Valeria Ferrari}
\email{ferrari@tandar.cnea.gov.ar}

\affiliation{Gerencia de Investigaci\'on y Aplicaciones, CNEA, Av. Gral Paz 1499 (1650) San Mart\'{\i}n, Pcia. Buenos Aires,
Argentina}

\begin{abstract}
Ab initio calculations using the local spin density approximation and also including the Hubbard $U$ have been performed 
for three low energy configurations of the interface between LaAlO$_3$ and TiO$_2$-anatase. Two types of interfaces have been
considered: LaO/TiO$_2$ and AlO$_2$/TiO, the latter with Ti-termination and therefore a missing oxygen. A slab-geometry calculation
was carried out and all the atoms were allowed to relax in the direction normal to the interface.

In all the cases considered, the interfacial Ti atom acquires a local magnetic moment and its formal valence is less than $+4$.
When there are oxygen vacancies, this valence decreases abruptly inside the anatase slab while in the LaO/TiO$_2$ interface the changes
are more gradual. 

\end{abstract}


\date{\today }
\maketitle

\section{Introduction}

Complex oxide heterostructures have been the subject of many recent papers,
both experimental and theoretical, as their interesting interface
properties promise to pave the way towards novel electronic devices. 
With the current available experimental techniques oxide thin films can be
produced with a high degree of cristallinity and the electronic
structure of their surfaces and interfaces can be precisely determined.
 The most studied system of this type has been  LaAlO$_3$/SrTiO$_3$
(LAO/STO), that consists of two perovskite structures stacked along the
(001) direction\cite{otrainterfaz}, in which they present alternate
layers of LaO, AlO2, TiO2 and SrO with a very small lattice mismatch
between them.
 A similar but less studied interface is that of LAO with TiO$_2$-anatase,
that presents an even smaller lattice mismatch, namely less than 0.1\%,
along the (001) direction \cite{kitazawa}. TiO2 is a key material for 
most applications, including catalytic and optical devices, sensors, 
optoelectronics and spintronics. Anatase thin films are frequently
grown over LAO by pulsed laser deposition. It seems therefore important
to perform a careful characterization, both from
theory and experiments, of the film/substrate interface. Just to mention one
example where this study may be relevant, room temperature ferromagnetism
has been obtained  from doped and undoped anatase films grown over LAO and 
the results strongly depend on the growth conditions \cite{cita}.

 There are two possible interfaces for the system LAO/TiO$_2$-anatase,
namely  LaO facing TiO$_2$ and AlO$_2$ facing TiO$_2$\cite{slafes,japoneses}.
 Due to the ionicity of the component oxides, and in particular, to the
fact that the layers of LAO (AlO$_2$ and LaO) have alternating formal
charges (-1 and +1, respectively), there is an interfacial formal excess
charge that should be compensated either by the presence of terraces with
different stacking, or by  oxygen vacancies or by atomic interdiffusion.

\begin{figure*}[htb]
\begin{center}
\includegraphics[width=1.0\textwidth,keepaspectratio]{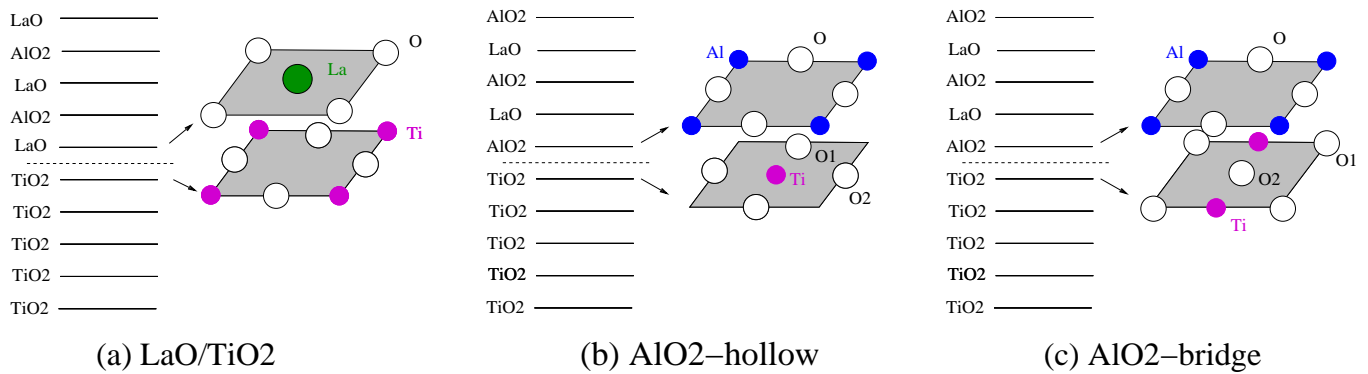}
\caption[]{Slab structures for three low energy configurations. (a) Interface
LaO/TiO$_2$; (b) Interface AlO$_2$/TiO, with Ti atom in the hollow position; (c) Interface 
AlO$_2$/TiO, with Ti atom in the bridge position. Oxygen atoms are co-planar with La and Al ions and 0.4 \AA off-plane in 
the Ti layers. O1 and O2 are below and above the Ti-plane, respectively. Oxygen O2 is closer to LAO and 
it is removed in cases (b) and (c).}.
\label{fig:1}
\end{center}
\end{figure*}

Experimental work \cite{japoneses} has shown that, in the case of 
LAO/TiO$_2$-anatase, terraces with both types of interfaces appear and approximately in the same 
proportion. For the theoretical modeling of this system the fact that the
TiO$_2$-anatase layers along the (001) direction are not strictly planar must be taken into
account.  As a consequence, the anatase interface can be either 
oxygen-terminated and thus neutral (with formal valence +4 for the interfacial Ti ion), 
or Ti-terminated. The latter situation is equivalent to considering 
surface oxygen vacancies and gives rise to a change in the valence of the 
interfacial Ti ion. 

In this paper we study the electronic and magnetic 
properties of interfaces for different stackings of LAO/TiO$_2$ using density functional theory\cite{DFT} in 
the local density approximation (LSDA) and also 
considering electronic correlations in the framework of the LSDA+U approximation.

\section {Method of calculation}

To simulate the interfaces, we use a periodically repeated slab-geometry with
 5 layers of each of the component materials and enough empty space between slabs 
so that they do not interact with each other (see Fig.\ref{fig:1}). The cell parameters 
in the plane of the interface are kept fixed to those of the experimental bulk LAO (3.79 \AA)
\cite{kitazawa} and 
the positions of all atoms are allowed to relax in the out of plane direction  
until the forces are smaller than 0.05 eV/~\AA.

As in our previous work\cite{slafes,saul}, we use an odd number of layers 
of LAO to avoid the formation of a large dipole moment, but we are 
aware that with this procedure it is difficult to assure whether  the interface  
will be conducting or not. The reason lies in the fact that an LSDA calculation for
a positively (negatively) charged system will locate the Fermi level in the conduction (valence) band.

We use the Wien2k code, that is an implementation of the full potential
linear augmented plane waves method (FP-LAPW)\cite{wien}. The 
calculations are 
scalar relativistic and the parameters used are listed on 
Ref.~\onlinecite{param}. It is well known that in the case of oxides there
is a band gap underestimation when using local exchange correlation functionals within 
density functional methods. For this reason, we have  performed 
calculations  with both the LSDA and the LSDA+U approximations\cite{LSDA+U}, 
using U=0.4 Ry for the Ti $d$ orbitals\cite{saul,ruben}, as in previous works. 
The comparison between the two procedures will evidence whether there is a qualitative 
difference in the interfacial properties due to electron 
correlations, as for example, a change in the valence of the Ti ions.

\section{Results}

The total energies of different possible structures (at 0K and 
with collinear spins) have been evaluated in previous works\cite{slafes,japoneses}, and in this
 paper we present the results for the lowest energy ones.
For the interface in which LaO faces TiO$_2$ (Fig.\ref{fig:1}(a)), the 
oxygen-termination for anatase is preferred. However, when 
AlO$_2$ faces TiO$_2$, the lower energy is for the Ti-termination, thus with 
oxygen vacancies\cite{slafes,japoneses}. 
Two types of Ti-terminated interfaces are studied: one where the interfacial Ti atom 
faces the hollow site of the Al atoms in the  AlO$_2$ layer (Fig.\ref{fig:1}(b)) and another 
one where the Ti 
atom faces a bridge position of the Al sites in the AlO$_2$ plane (Fig.\ref{fig:1}(c)).

In all the cases considered there is a 
total formal charge of +1 arising in (a) from the LAO slab while in cases (b) and (c) 
it is due to the contribution of both the LAO slab and the oxygen vacancy in the anatase slab. 
We point out that there are more possible structures for each type of stacking, and that
in this work we consider three of the lowest energy ones, those
that require more energy to separate LAO from anatase. 
LaO/TiO$_2$ has the
lowest energy, followed by AlO$_2$/TiO where  oxygen vacancies are present\cite{note} (i.e. O2 is removed in Fig.\ref{fig:1}(b) 
and (c)).
In the hollow structure one oxygen atom from the AlO$_2$ layer moves towards the interfacial Ti thus increasing its 
number of neighbors. This relaxation decreases the total energy of the 
system with respect to the bridge structure by 0.2 eV in LSDA and 0.4 eV in LSDA+U.

\begin{figure*}[htb]
\begin{center}$
\begin{array}{cc}
\includegraphics[width=0.36\textwidth,keepaspectratio,angle=-90]{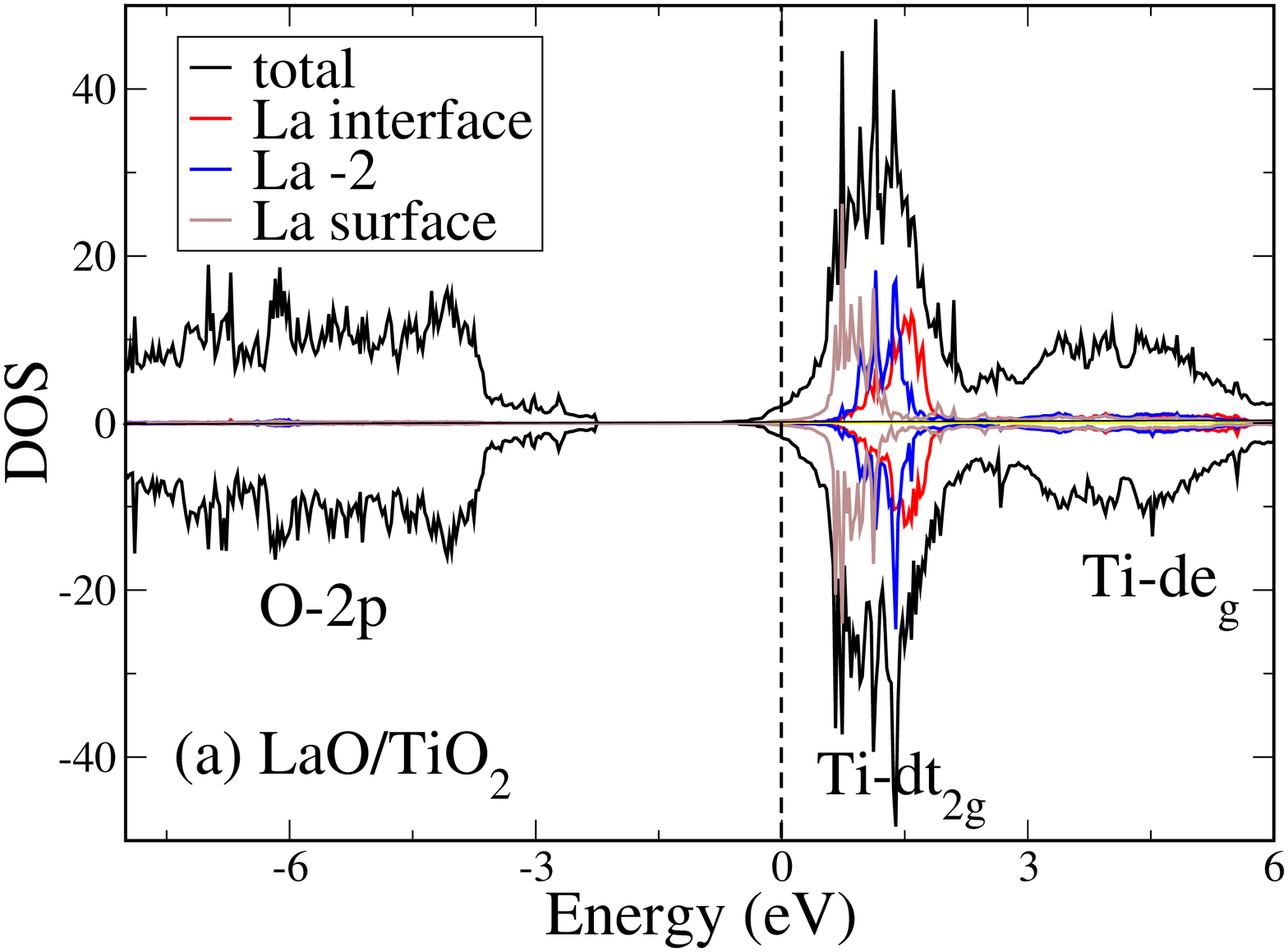}&
\includegraphics[width=0.36\textwidth,keepaspectratio,angle=-90]{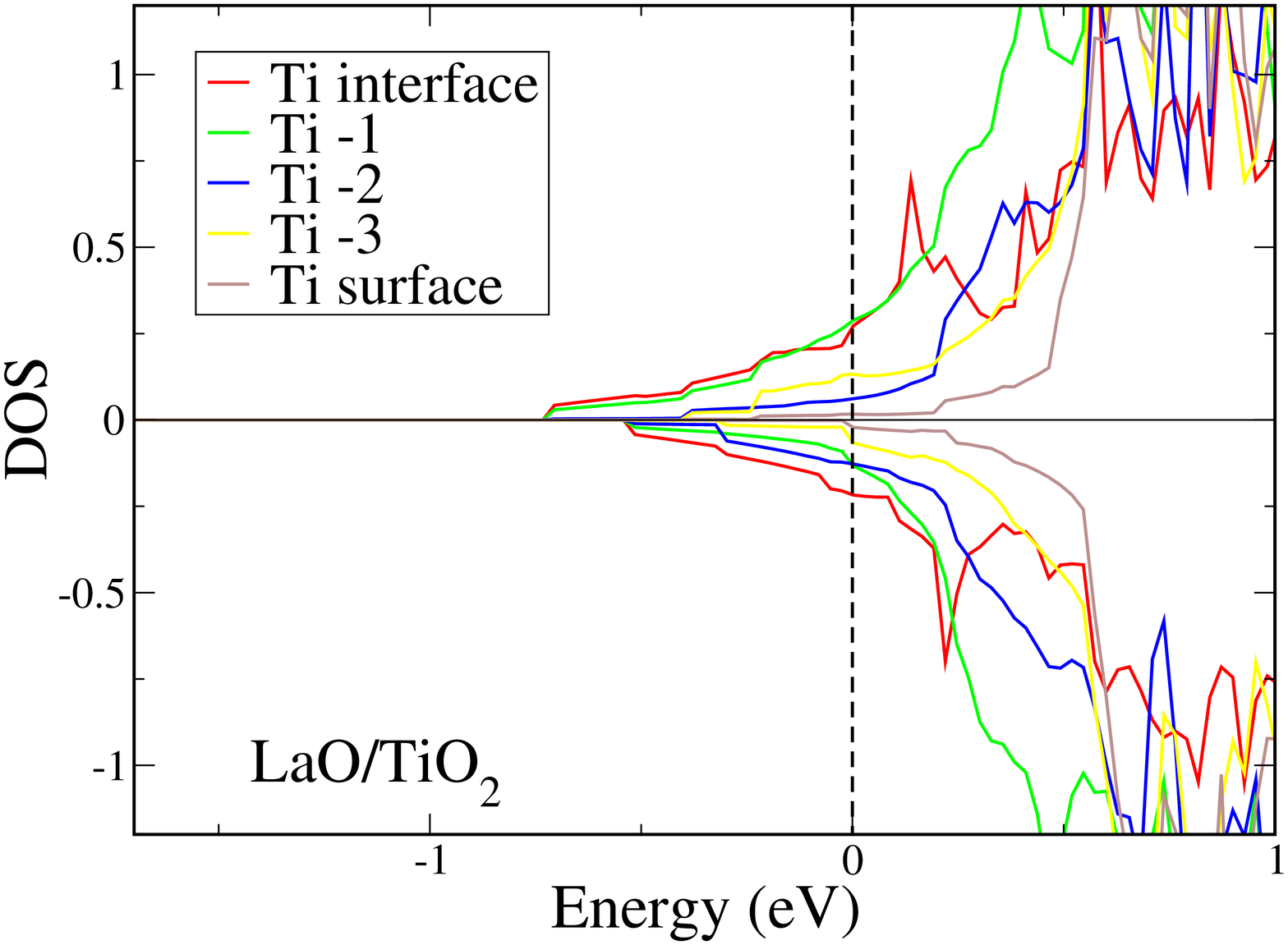}\\
\includegraphics[width=0.36\textwidth,keepaspectratio,angle=-90]{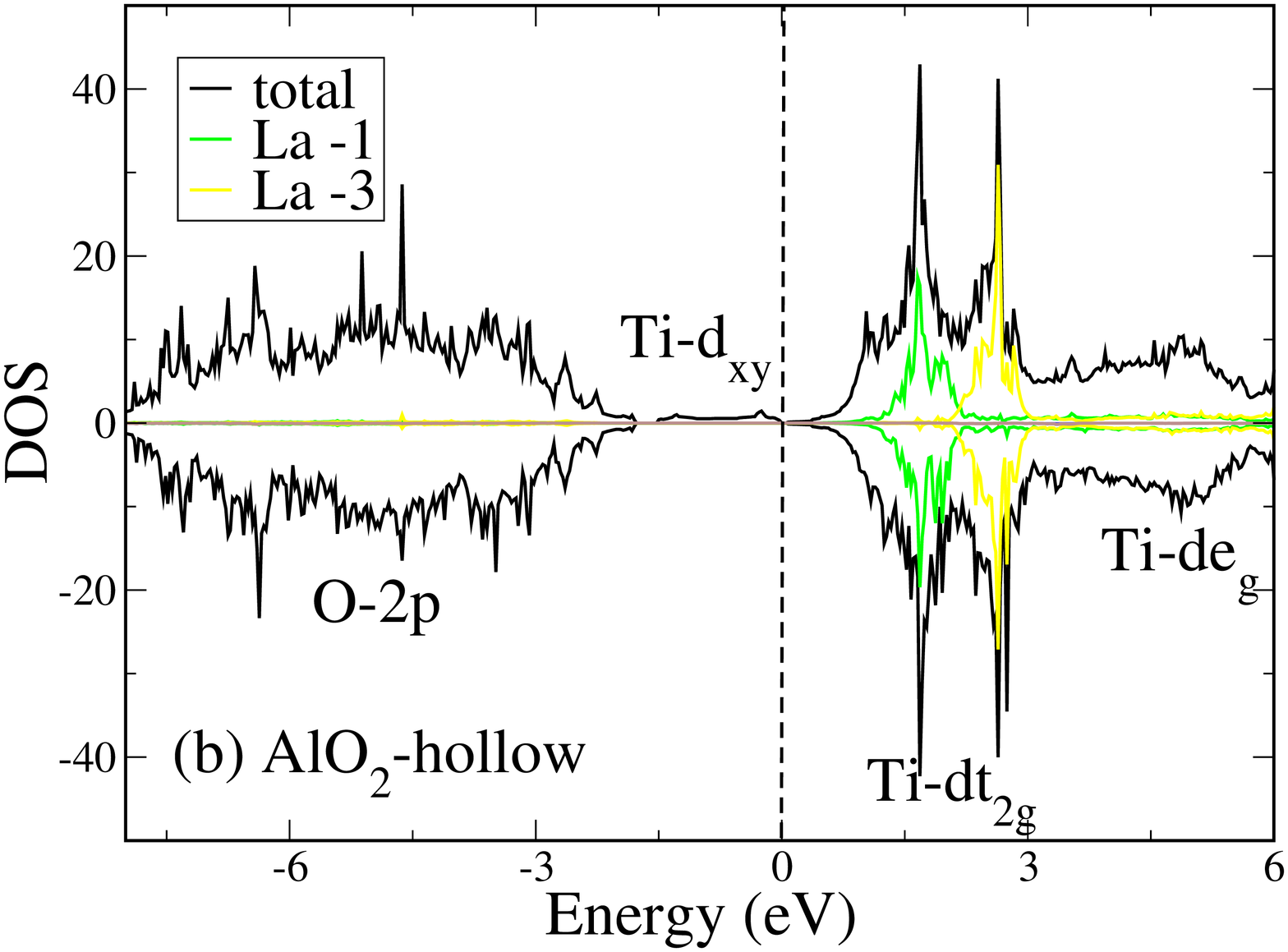}&
\includegraphics[width=0.36\textwidth,keepaspectratio,angle=-90]{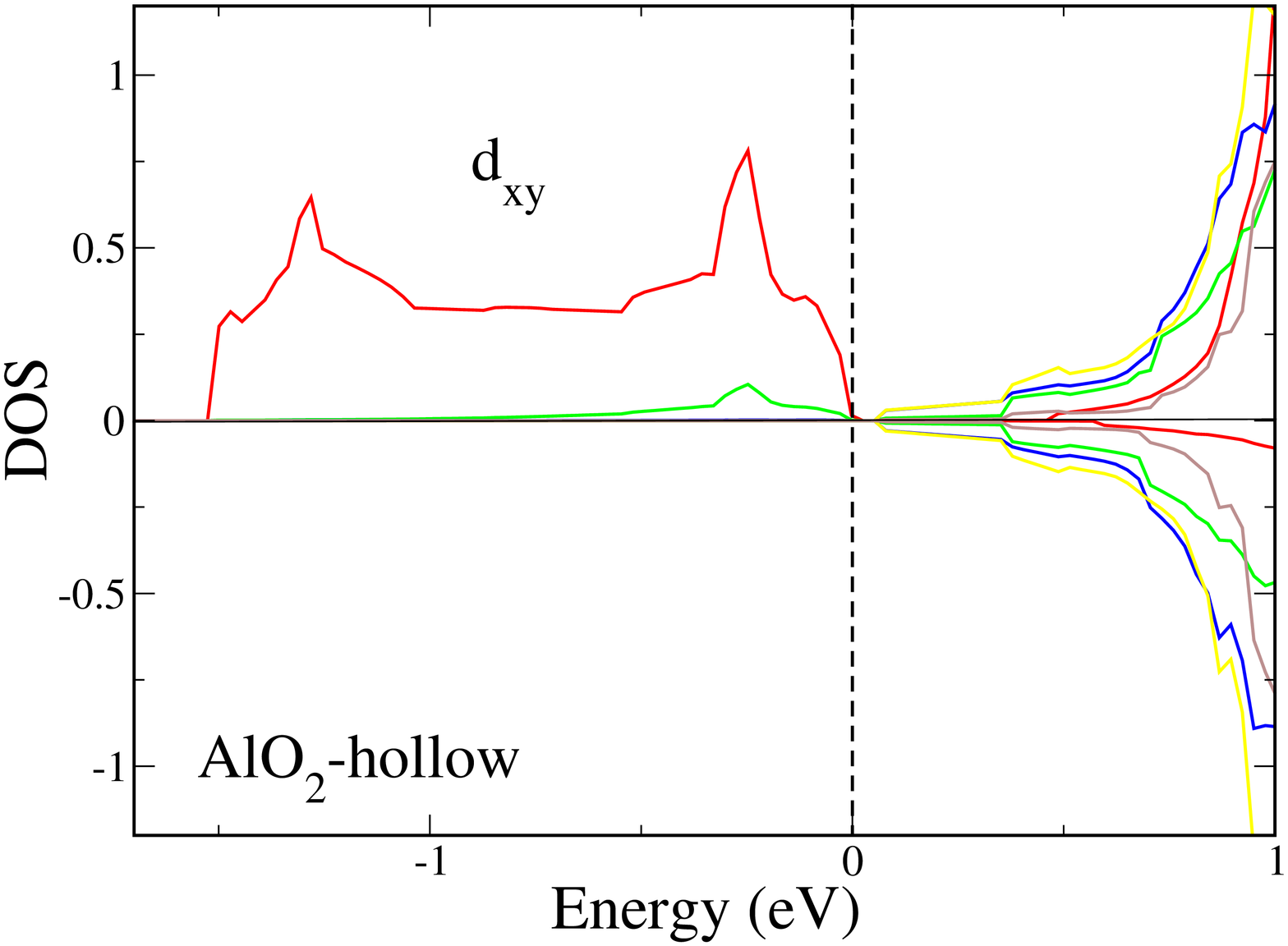}\\
\includegraphics[width=0.36\textwidth,keepaspectratio,angle=-90]{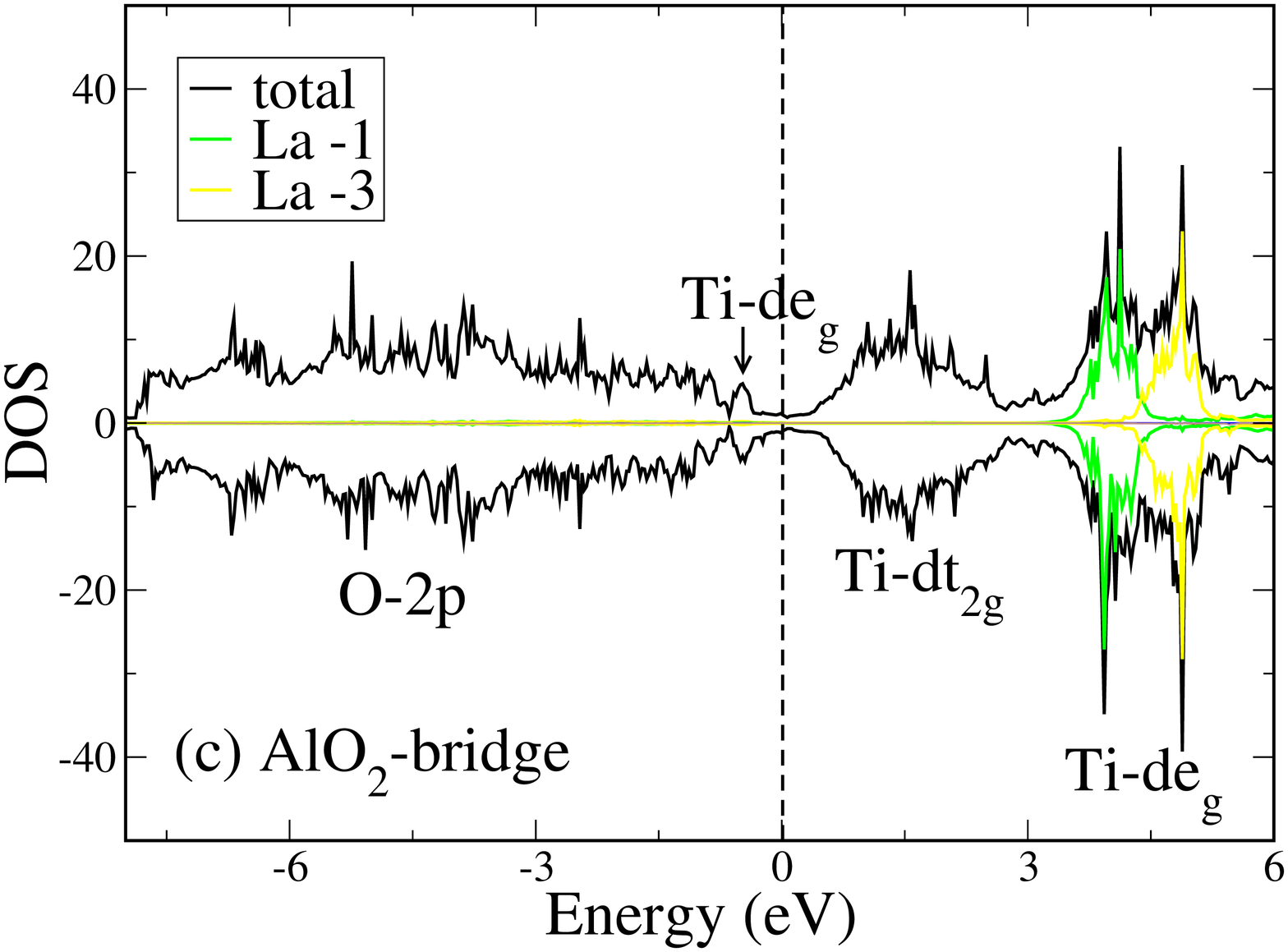}&
\includegraphics[width=0.36\textwidth,keepaspectratio,angle=-90]{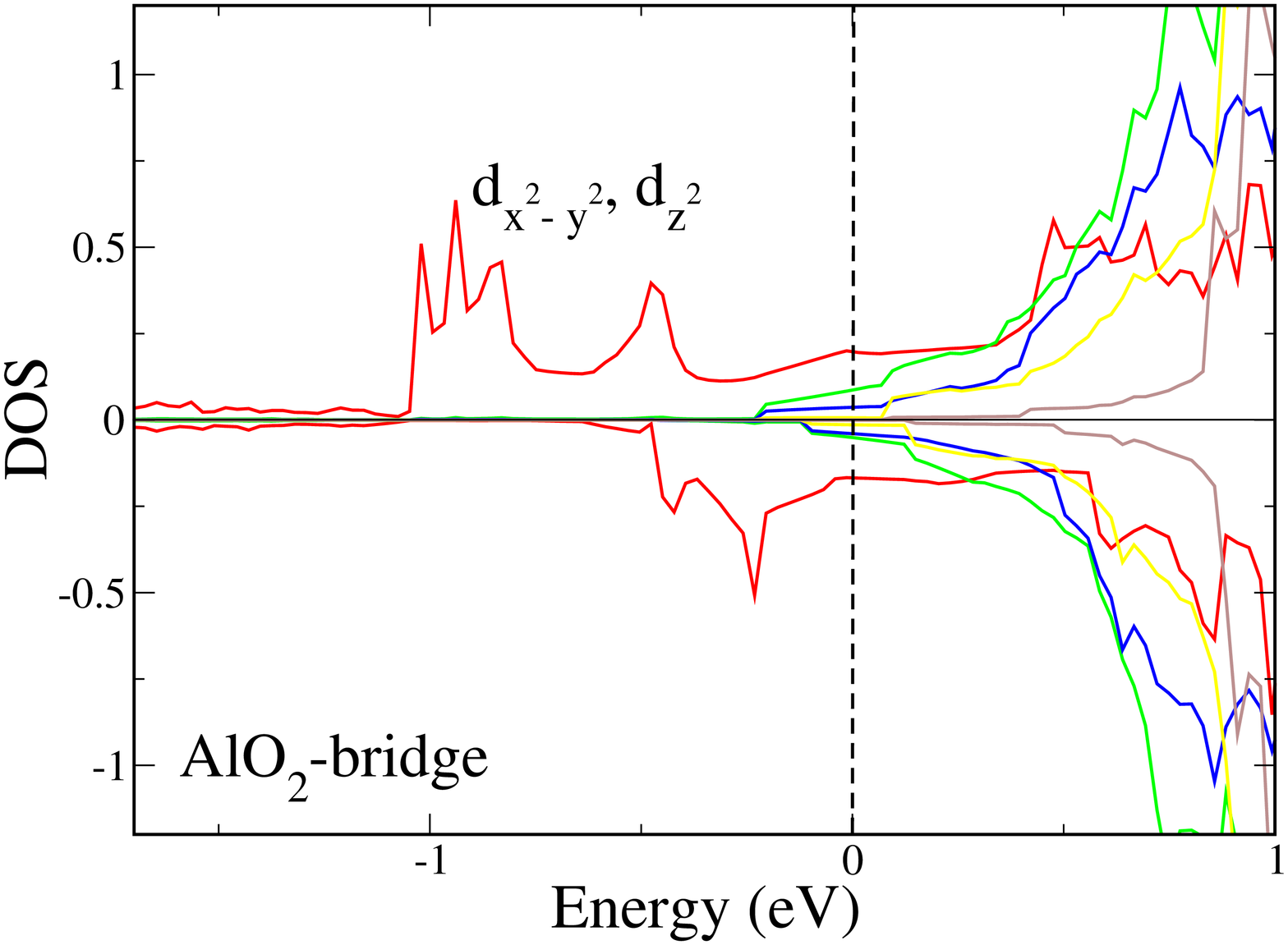}
\end{array}$
\caption{Densities of states for the three 
slabs of Fig. \ref{fig:1} in the LSDA+U approximation. (a) Top: LaO/TiO$_2$, (b) Middle: AlO$_2$-hollow, (c) Bottom: AlO$_2$-bridge.
The Fermi level (E$_f$) is marked by the dotted vertical line. Positive and negative values correspond 
to majority and minority spins, respectively.
Left: total densities of states of the slabs and La levels. Right: Partial DOS on the Ti atoms zoomed-in around E$_f$. The labels of the Ti projected DOS are the same for the 3 cases.
In the legend, the number by the atoms indicate the number of planes away from the interface.}
\label{fig:2}
\end{center}
\end{figure*}

Fig.\ref{fig:2}  shows the densities of states (DOS) for the three structures
depicted in Fig.\ref{fig:1}. We present the results for the LSDA+U approximation as they
are similar to the LSDA ones and if there are any differences, they are mentioned in the text. On the
left side of Fig.\ref{fig:2}, we present  the total DOS and  the
partial contributions from the La atoms. On the right side, the DOS projected on the Ti atoms are shown
in a zoomed-in region close to the Fermi energy (E$_f$).  
As in the case of pure anatase, the valence band is mostly from the oxygen atoms and the 
conduction band from the Ti atoms, with the $d$ orbitals clearly separated 
 by symmetry: the $t_2g$ levels closer to  E$_f$ and the 
e$_g$ ones at higher energies. There are no Al states close to E$_f$, but
oxygen atoms from LAO contribute significantly to the higher energy part of
the valence band.

\begin{figure*}[htb]
\begin{center}$
\begin{array}{ccc}
\includegraphics[width=0.3\textwidth,keepaspectratio]{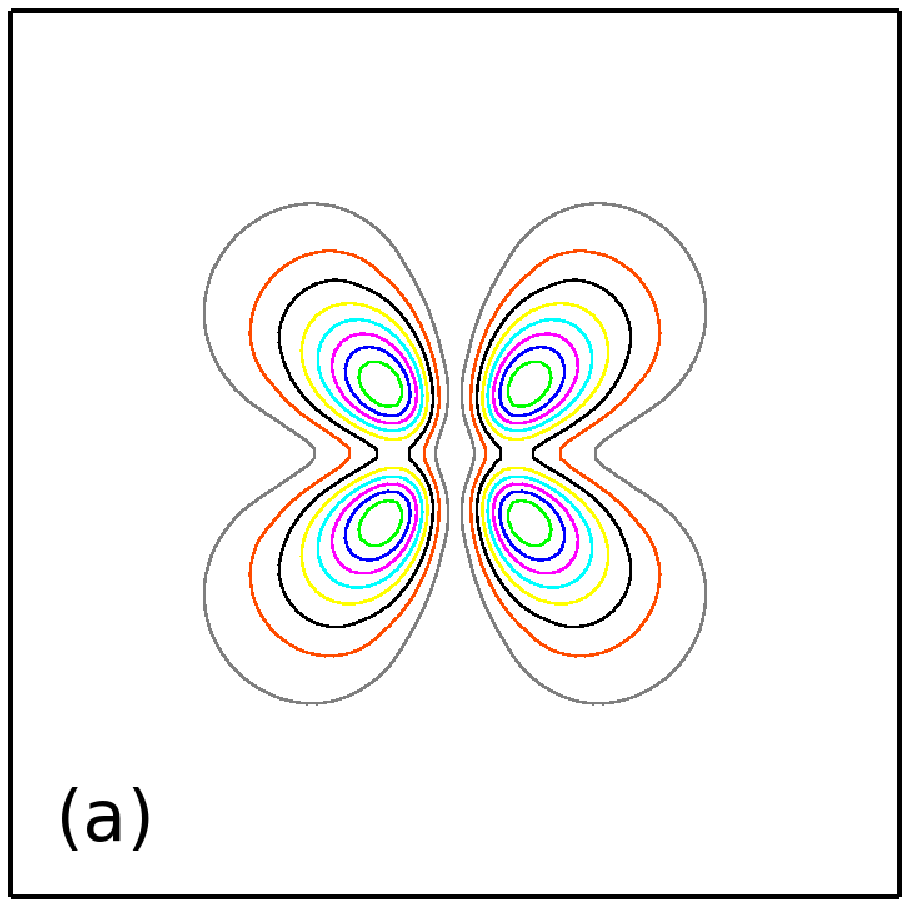}&
\includegraphics[width=0.3\textwidth,keepaspectratio]{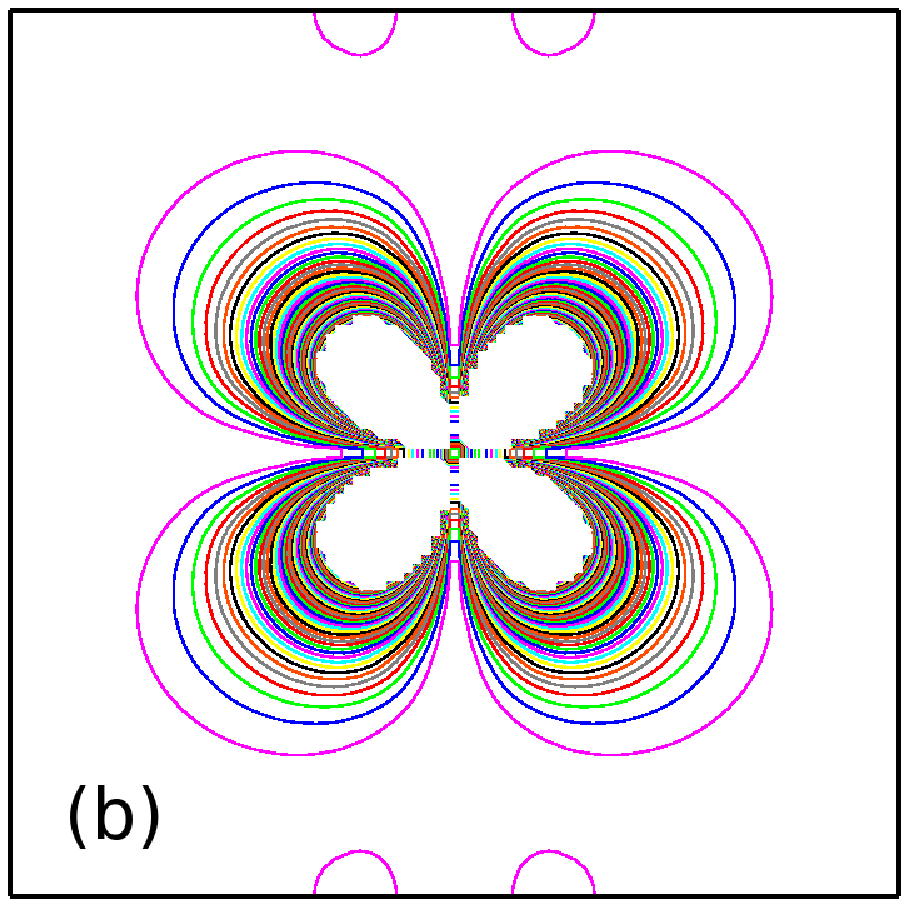}&
\includegraphics[width=0.3\textwidth,keepaspectratio]{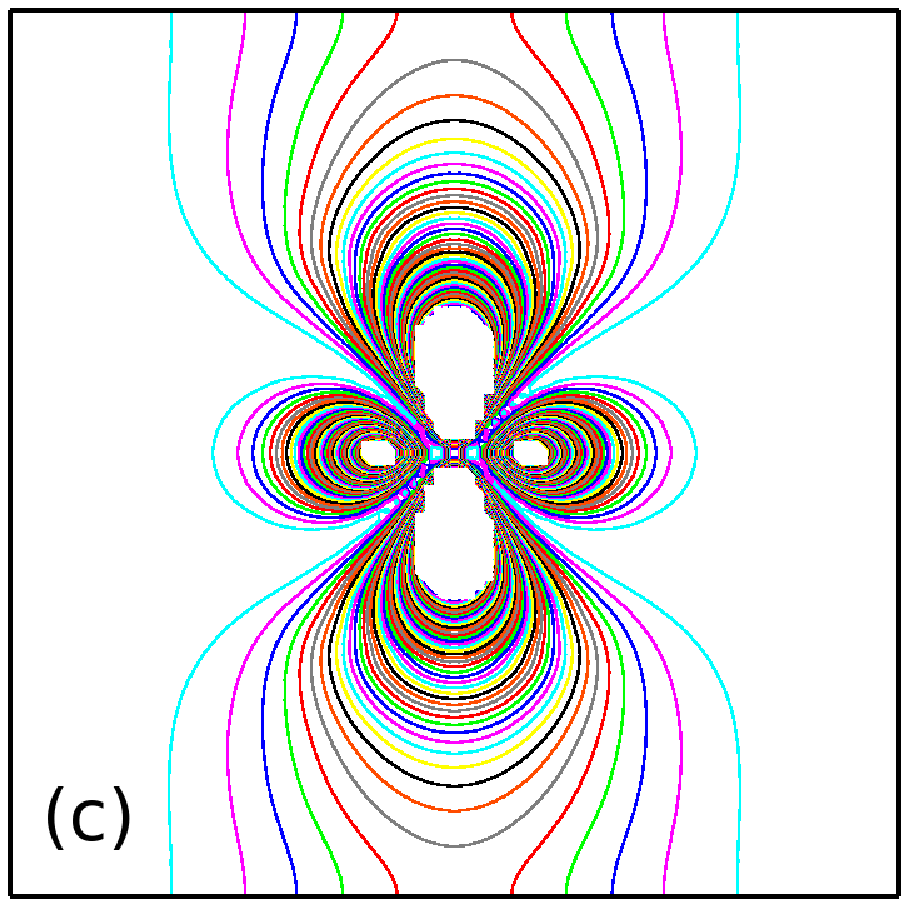}
\end{array}$
\caption[]{LSDA+U charge density of states close to E$_f$ in the plane defined by the Ti-Ti bonds 
for one unit cell. The occupied states in the energy range: [E$_f$-1.5eV, E$_f$] are plotted
for the three structures in Fig.\ref{fig:1}, namely (a) LaO/TiO$_2$, (b) AlO$_2$-hollow, (c) AlO$_2$-bridge. Charge isolines are spaced 0.01, from 0.01 to 1.00. In each case, the Ti interfacial atom is placed at the center of the figure, so as 
to make them comparable.}
\label{fig:3}
\end{center}
\end{figure*}

\begin{figure}[htb]
\begin{center}
\includegraphics[width=0.4\textwidth,keepaspectratio,angle=-90]{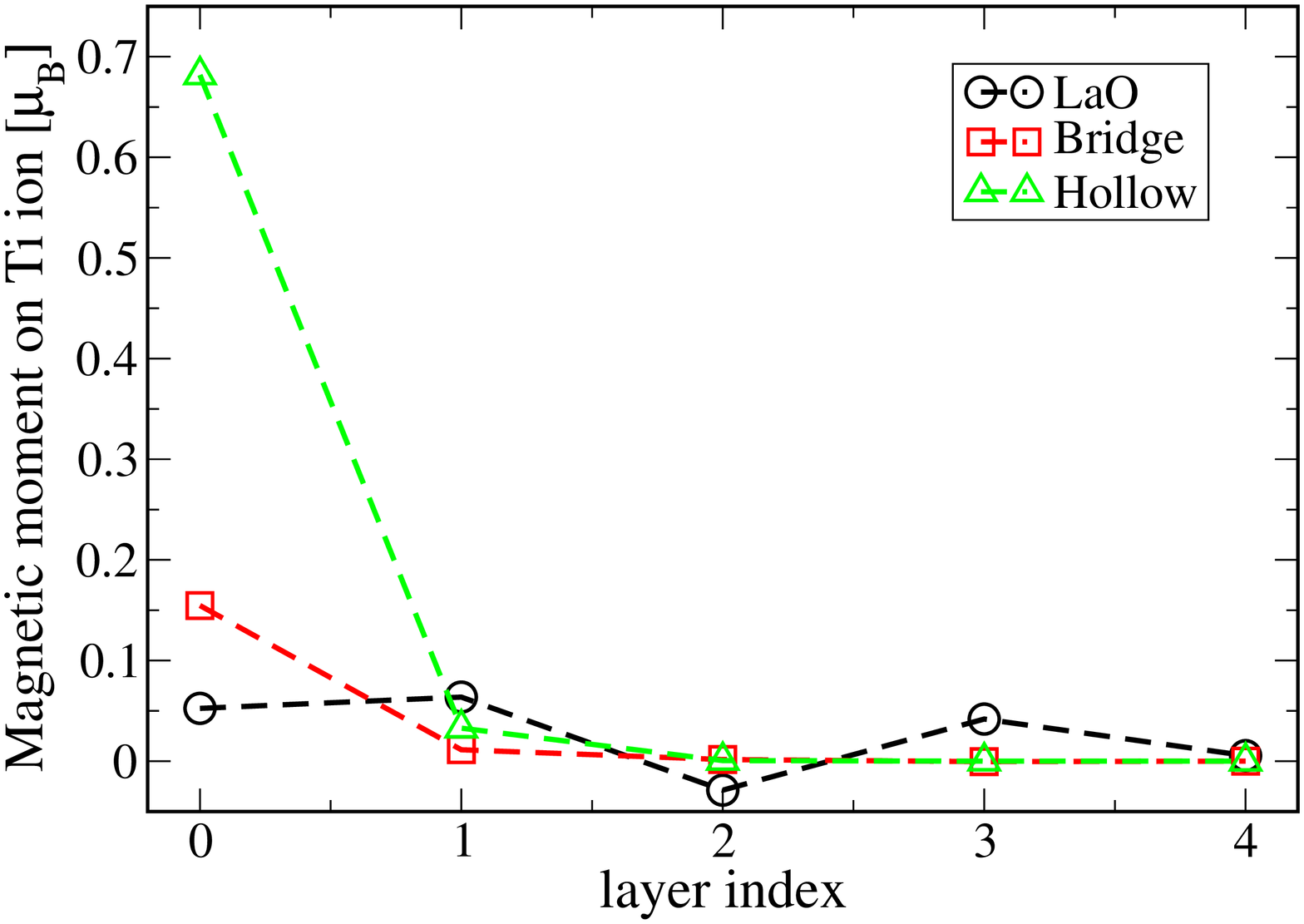}
\caption[] {Magnetic moment inside the Ti muffin-tin spheres
as a function of the penetration in the anatase slab. The interface is at layer 0 and
the surface is at layer 4.
}
\label{fig:4}
\end{center}
\end{figure}

\vspace*{2mm}
\noindent{\bf LaO/TiO$_2$ interface}
\vspace*{2mm}

In this interface (Fig.\ref{fig:1}(a)), E$_f$ is in the conduction band, as expected due to 
the formal charge in the LAO slab. The largest contribution to the occupied states close to E$_f$
comes from the interfacial Ti ion and  the rest of the Ti atoms contribute less as the distance to the interface increases, as shown in Fig.\ref{fig:2}(top-right). This is true for both
LSDA and LSDA+U approximations.  
It is interesting to note that similar layer projected DOS were obtained for LAO/STO \cite{Karolina} even though in that structure, the TiO$_2$ planes alternate with 
SrO layers. The authors also find that the charge distribution is not affected by the inclusion of correlation.
There is certainly no contribution from the
interfacial La ions to the occupied states, as that peak is shifted towards the
right when compared with the La atom at the free surface of the slab (see Fig.\ref{fig:2}(a)).
The charge density at the interface is plotted in Fig.\ref{fig:3}(a) and
shows clearly the shape of the Ti $d$ orbitals. The main occupation comes from
d$_{xz}$ and d$_{3z2-r2}$.
A small total magnetic moment is obtained for the unit cell, distributed among all the 
Ti atoms (see Fig.\ref{fig:4}).

\vspace*{2mm}
\noindent{\bf AlO$_2$/TiO-hollow interface}
\vspace*{2mm}

This interface (Fig.\ref{fig:1}(b)), presents a difference between LSDA and LSDA+U approximations.
In LSDA+U, the majority spin of the interfacial Ti d$_{xy}$ orbital shifts towards lower
energies, so that the system becomes a 
semiconductor with a small gap of 0.1 eV (see Fig.\ref{fig:2}(b)) This shift is smaller in the 
LSDA, so that the system is metallic. 
In Fig.\ref{fig:3}(b) we show the charge density at the anatase side of the interface where
the $d_{xy}$ symmetry is evident.
Both the charge and the magnetic moment at the Ti ions are large at the interface and
decrease abruptly inside the anatase slab (see Fig.\ref{fig:4}).

\vspace*{2mm}
\noindent{\bf AlO$_2$/TiO-bridge interface}
\vspace*{2mm}

In this interface (Fig.\ref{fig:1}(c)), E$_f$ lies in the conduction band, and the La peaks are further
away from it, overlapping the $t_{2g}$ orbitals of the Ti atoms. The
same result is  obtained with LSDA.
The interfacial Ti orbitals have different symmetry from those of the hollow case, being of $e_g$ character:
$d_{x2-y2}$ and $d_{3z2-r2}$, as can be seen in Fig.\ref{fig:2} and Fig.\ref{fig:3}. In this case, 
the $e_g$ levels split into two regions: one occupied band close to E$_f$
and another one at energies above the $t_{2g}$ orbitals, in the conduction band.
The extra charge and the magnetic moment are both localized at the interface, near
the oxygen vacancy, and their magnitude decreases abruptly inside the anatase slab as can
be seen in Fig.\ref{fig:4}. In this case, the interfacial Ti formal valence  and
the magnetic moment are smaller than in the hollow structure.

 It is interesting
to note that in  AlO$_2$/TiO interfaces, there is no contribution of the 
Ti atom at the free surface in the occupied region of the conduction band, showing that a slab
with 5 layers is a good enough approximation to account for interface properties.

There has been some controversy concerning the appearance of interfacial
ferromagnetism in the LAO/STO interface \cite{ivan}, which  may also
be the case here. For this reason we performed calculations with a
double size unit cell (54 atoms), for the AlO$_2$/TiO interfaces which present localized 
magnetism at the interface. We studied the magnetic interaction between the 
interfacial Ti atoms within the LSDA+U method, setting them both
in  parallel and antiparallel configurations. The two solutions exist but 
the antiparallel one is lower in energy, indicating that a long-range ferromagnetic interaction
will not arise from the considered concentration and distribution of oxygen vacancies in  AlO$_2$/TiO interfaces.
We do not exclude the possibility of ferromagnetism in LaO/TiO$_2$, but it 
would be minor and
due to fact that all the Ti atoms have local
magnetic moments, many magnetic configurations should be considered for this study.

To explore the consequences of a lower vacancy concentration, and  
also of a formally neutral system, we considered the double size unit cell 
in the hollow configuration with only one interfacial oxygen vacancy. 
In this calculation none of the Ti ions resulted Ti+3 and  
no magnetic moments appeared.
Thus, a charge imbalance is needed to obtain Ti+3 ions and  is also
a necessary (but not sufficient) condition for the appearance of 
magnetism.

\section{Conclusions}

The principal conclusion of this work is that whatever the origin of the charge at the interfaces, its main effect is 
to change the valence of the Ti atoms, either if it is due to the layered-structure of LAO or to the presence
of oxygen vacancies in anatase.

In AlO$_2$/TiO, there are interfacial oxygen vacancies and the  interfacial
Ti atom close to the vacancy position, acquires an extra charge. In LaO/TiO$_2$, there are no oxygen vacancies as they are not 
electrostatically favored  and the extra charge is distributed among all Ti atoms, with decreasing value
as the distance from the interface increases.  
Our results indicate that the LaO/TiO$_2$ interface 
spans through several anatase layers  while the AlO$_2$/TiO one is more localized and presents
large local magnetic moments.

In experimental samples, co-existence of the studied interfaces and possibly others as well, is expected. As a
consequence, small patches with different magnetic order might appear \cite{natphys}, thus giving rise 
to sample-dependent results.

\section {Acknowledgments}

We thank R. Weht for a careful reading of the manuscript. M.W. and V.F. are members of CONICET-Argentina and acknowledge 
support by grant PIP-CONICET00038.


\begin{thebibliography}{99}
\bibitem{param} For the FP-LAPW calculations we use the 
following parameters:
Atomic sphere radii of 1.7 Bohr for Ti and Al, 1.4 Bohr for O and 2.5 for La.
The convergence control parameter RK$_{max}$ is chosen to be 6, that
corresponds to an energy cutoff of 250 eV.
The employed vacuum space between slabs is 20 \AA.
\bibitem{note} In this work the number of vacancies considered is large, but the simulation of smaller
vacancy concentrations would imply, due to the anatase structure, to
start from corrugated interfaces that may be the the subject of future work.
\end{thebibliography}
\end{document}